\documentclass[12pt]{article}
\begin{document}
\begin{center}
\thispagestyle{empty}
 \vspace{2cm}
\textbf{\large The thermal waves induced by ultra-short laser
pulses in $n$-dimensional space-time}

\bigskip\bigskip
Janina Marciak-Kozlowska\footnote{Corresponding author}\\
Miroslaw Kozlowski

\bigskip\bigskip
Institute of Electron Technology, Al. Lotnik\'{o}w 32/46, 02-668
Warsaw, Poland
\end{center}

\vspace{2cm}
\begin{abstract}
In this paper the heat waves, induced by ultra-short laser pulses
are considered. The hyperbolic heat transport in $n$-dimensional
space-time is formulated and solved. It is shown that only for
$n-$odd for heat waves the Huygens principle is fulfilled. The
heat transport experiment for Cu$_3$Au alloy is considered.
\end{abstract}

\newpage
\section{Introduction}
The fact that we perceive the world to have three spatial dimensions is something so familiar to our experience of its
structure that we seldom pause to consider the direct influence this special property has upon the laws of physics.
Yet some have done so and there have been many intriguing attempts to deduce the expediency or inevitability of a
three-dimensional world from the general structure of the physical law themselves.

As earlier as 1917 P. Ehrenfest~\cite{1} pointed out that neither
classical atoms nor planetary orbits can be stable in a space with
$n>3$ and traditional quantum atoms cannot be stable
either~\cite{2}. As far as $n<3$ is concerned, it has been
argued~\cite{3} that organism would face insurmountable
topological problem if $n=2$: for instance two nerves cannot
across. In the following we will conjecture that since $n=2$
offers vastly less complexity that $n=3$, worlds with $n<3$ are
just too simple and barren to contain observers. Since our
Universe appears governed by the propagation of classical and
quantum waves it is interesting elucidate the nature of the
connection the properties of the wave equation and the spatial
dimensions.

In this paper we describe the partial differential equation (PDE) for the propagation of the thermal waves in
$n-$dimensional space time. It is well known that for heat transport induced by ultra-short laser pulses
(shorter than the relaxation time) the governing equation can be written as~\cite{4}
    \begin{equation}
    \frac{1}{v^2}\frac{\partial^2 T}{\partial t^2}+\frac{1}{D}\frac{\partial T}{\partial
    t}+\frac{2Vm}{\hbar^2}T=\nabla^2T,\label{eq1}
    \end{equation}
   where $T$ is the temperature, $v$ denotes the thermal wave propagation, $m$ is the mass of heat carriers
   and $V$ is the potential.

In monograph~\cite{4} the solution of the equation for
one-dimensional case, $n=1$ was obtained. In this paper we
develope and solve the analog of the equation for $n=$ natural
numbers $n=1,2,\cdots,$ separately for $n=$ odd and $n=$ even. The
Huygens' principle for thermal wave will be discussed. It will be
shown that for thermal waves only in odd dimensional space the
waves propagate at exactly a fixed space velocity $v$ without
``echoes'' assuming the absence of walls (potentials) or
inhomogeneities.

The three-dimensional heat transfer induced by ultra-short laser
pulses in Cu$_3$Au alloy will be suggested.

\section{The master equation for the thermal waves in $n-$dimensions}
In the following we consider the $n-$dimensional heat transfer phenomena described by the equation~\cite{4}
    \begin{equation}
    \frac{1}{v^2}\frac{\partial^2 T}{\partial t^2}+\frac{1}{D}\frac{\partial T}{\partial t}+\frac{2Vm}{\hbar^2}T=
    \nabla^2 T\label{eq2}
    \end{equation}
where temperature $T$ is the function in the $n-$dimensional space
    \begin{equation}
    T=T(x_1, \ldots,x_n,t)\label{eq3}
    \end{equation}
    We seek solution of equation~(\ref{eq2}) in the form:
    \begin{equation}
    T(x_1, x_2,\dots,x_n)=e^{-\frac{t}{2\tau}}u(x_1, \dots, x_n,t)\label{eq4}
    \end{equation}
    After substitution of Eq.~(\ref{eq4}) to Eq.~(\ref{eq2}) one obtains
    \begin{equation}
    \frac{1}{v^2}\frac{\partial^2 u}{\partial t^2}-\nabla^2u+qu=0\label{eq5}
    \end{equation}
    where
    $$
    q=\frac{2Vm}{\hbar^2}-\left(\frac{mv}{2\hbar}\right)^2
    $$
    for $ D=\frac{\hbar}{m}$~\cite{4}.

 We can define the distortionless thermal wave as the wave which preserves the shape in the field
 of the potential $V$. The condition for conserving the shape can be formulated as
    \begin{equation}
    q=\frac{2Vm}{\hbar^2}-\left(\frac{mv}{2\hbar}\right)^2=0\label{eq7}
    \end{equation}
When Eq.~(\ref{eq7}) holds Eq.~(\ref{eq5}) has the form
    \begin{equation}
    \frac{1}{v^2}\frac{\partial^2 u}{\partial t^2}-\nabla^2 u=0\label{eq8}
    \end{equation}
and condition~(\ref{eq7}) can be written as
    \begin{equation}
    V\tau\sim\hbar\label{eq9}
    \end{equation}
We conclude that in the presence of the potential  energy $V$ one
can observe the undisturbed thermal wave only when the Heisenberg
uncertainty relation~(\ref{eq9}) is fulfilled.

The solution of the Eq.~(\ref{eq8}) for the $n-$odd can be find
in~\cite{5}. First of all let us change the variables in
Eq.~(\ref{eq8})
    $$
    v:     t\to t',   \quad x\to x', \quad u\to u'
    $$
and obtain
    \begin{equation}
    \frac{\partial^2 u'}{\partial t'^2}-\nabla^2u'=0.\label{eq10}
    \end{equation}
For
    \begin{eqnarray}
    \lim_{x',t'\to(x^0,0)} u'(x',t')&=&g(x_0),\label{eq11}\\
    \lim_{x',t'\to(x^0,0)}\frac{\partial u(x',t')}{\partial t'}&=&h(x_0),\nonumber
    \end{eqnarray}
the solution have the form~\cite{5}
    \begin{eqnarray}
    u'(x',t')&=&\frac{1}{\gamma_n}\left[\left(\frac{\partial}{\partial t'}\right)
\left(\frac{1}{t'}\frac{\partial}{\partial
t'}\right)^{\frac{n-3}{2}}\left(t'^{n-2}
\oint_{\partial B(x',t')}gdS\right)\right.\nonumber\\
&&\mbox{}+\left.\left(\frac{1}{t'}\frac{\partial} {\partial
t'}\right)^{\frac{n-3}{2}}\left(t'^{n-2}\oint_{\partial
B(x',t')}hdS\right)\right]\label{eq12}
    \end{eqnarray}
and $\gamma_n=1 \cdot 3 \cdot 5 \cdots (n-2)$.

For $n-$ even the solution of equation~(\ref{eq10}) have the form~\cite{5}.
\begin{eqnarray}
    u'(x',t')&=&\frac{1}{\gamma_n}\left[\left(\frac{\partial}{\partial t'}\right)
\left(\frac{1}{t'}\frac{\partial}{\partial
t'}\right)^{\frac{n-2}{2}}\left(t'^{n}
\oint_{B(x',t')}\frac{g(y')dy'}{(t'^2-|y'-x'|^2)^{\frac{1}{2}}}\right)\right.\nonumber\\
&&\mbox{}+\left.\left(\frac{1}{t'}
\frac{\partial}{\partial t'}\right)^{\frac{n-2}{2}}\left(t'^{n}\oint_{B(x',t')}
\frac{h(y')dy'}{(t'^2-|y'-x'|^2)^{\frac{1}{2}}}
\right)\right]\label{eq13}
    \end{eqnarray}
    and $\gamma_n=2\cdot4\cdots(n-2)\cdot n$. In formulae
    (\ref{eq12}) and (\ref{eq13}) $\oint$ denotes integral over
    $n-$space.

    Considering formulae~(\ref{eq12}) and (\ref{eq13}) we conclude that for $n-$odd the solution~(\ref{eq12})
    is dependent on the value of functions $h$ and $g$ only on the hypersphere $\partial B(x', t')$. On the
    other hand for $n-$even the solution~(\ref{eq13}) is dependent on the values of the functions $h$ and $g$
    on the full hyperball $B(x',t')$. In the other words for $n-$odd $n\geq3$ the value of the initial
    functions $h$  and $g$ influence the solution~(\ref{eq13}) only on the surface of  the
    cone $\{(y',t'), t'>0, |x'-y'|=t'\}$. For $n=$ even the value of the functions $g$ and $h$
    influences the solution on the full cone. It means that the thermal wave induced by the
    disturbance for $n=$ odd have the well defined front. For $n-$even the wave influences space
    after the transmission of the front. This means that Huygens' principle is false for $n-$even.
    In conclusion: if we solve the wave equation in $n-$dimensions the signals propagate sharply
    (i.e. Huygens' principle is valid) only for dimensions $n=3, 5, 7,\ldots$. Thus three is the
    ``best of all possible'' dimensions, the smallest dimension in which signals propagate sharply.
\section{Suggestions for experimentalists}
The hyperbolic transport equation for heat transport~(\ref{eq1})
    \begin{equation}
    \frac{1}{v_T^2}\frac{\partial^2 T}{\partial t^2}+\frac{1}{D_T}\frac{\partial  T}{\partial t}+\frac{2Vm}{\hbar^2}T=
    \nabla^2T\label{eq14}
    \end{equation}
or mass transport
    \begin{equation}
    \frac{1}{v_{\rho}^2}\frac{\partial^2\rho}{\partial t^2}+\frac{1}{D_{\rho}}
\frac{\partial \rho}{\partial t}+\frac{2Vm}{\hbar^2}\rho=\nabla^2\rho\label{eq15}
    \end{equation}
are the damped wave equations. For very short time period $\Delta t\sim\tau$ both equations~(\ref{eq14})
and (\ref{eq15}) can be written  as
    \begin{equation}
    \frac{1}{v_{\rho,T}^2}\frac{\partial^2u_{\rho, T}}{\partial t^2}-\nabla^2u_{\rho, t}=0\label{eq16}
    \end{equation}
Eq.~(\ref{eq16}) is the generalization of equation~(\ref{eq8}).
The solution of equation~(\ref{eq16}) in $n-$ dimensional cases
are described by formulae (\ref{eq12}) and (\ref{eq13}).

As was discussed in paragraph~2 only in $3-$dimensional case the
Huygens principle is fulfilled. It seems that in order to observe
the thermal wave not disturbed by the "echoes" and with sharp
front \textit{the true three-dimensional experiment} must be
performed. Moreover the experiment must be performed in the
relaxation regime, i.e. for materials with relatively long
relaxation time. The best candidates for ``relaxation materials''
it the Cu$_3$Au alloy~\cite{6}. As was shown in paper~\cite{6} the
relaxation time is of the order of $10^4$~s in the temperature
range 650--660 K. For $t>660$~K  the abrupt increasing, up to
$1.5\cdot10^5$~s (due to order $\to$ disorder transition) was
observed.


\newpage
\begin{thebibliography}{99}
\bibitem{1}P. Ehrenfest, \textit{Proc. Amsterdam Acad.}, \textbf{20},  (1917), p.~200.
\bibitem{2}F.R. Tangherlini, \textit{Nuovo Cimento}, \textbf{27}, (1963), p.~636.
\bibitem{3}C. J. Whitrow, \textit{Brit. J. Phil.}, \textbf{6}, (1955),   p.~13.
\bibitem{4}M. Kozlowski, J. Marciak-Kozlowska, \textit{From quarks to bulk matter}, \textit{Hadronic Press}, USA, 2001.
\bibitem{5}L. C. Evans, \textit{Partial Differential Equations}, American Mathematical Society, USA, 1998.
\bibitem{6}T. Hashimoto et al., \textit{Phys. Rev.} \textbf{B13}, (1997), p.~1119.
\end{thebibliography}
\end{document}